\documentclass{article}
\usepackage{spconf,amsmath,epsfig,hyperref}

\title{Towards Physically-consistent, Data-driven Models of Convection}
\name{Tom Beucler$^{1,2} $\thanks{Thanks to NSF grants AGS-1734164 and OAC-1835863 for funding, and to Paul O'Gorman and David Randall for conversations that helped formulate  our physical rescalings (Section 3).}, Michael Pritchard$^{1} $, Pierre Gentine$^{2} $, Stephan Rasp$^{3} $}
\address{$^{1} $Department of Earth System Science, University of California, Irvine, CA, USA,\\
$^{2} $Department of Earth and Environmental Engineering, Columbia University, New York, NY, USA,\\
$^{3} $Technical University of Munich, Munich, Germany}

\begin{document}
%
\maketitle

\begin{abstract}
Data-driven algorithms, in particular neural networks, can emulate the effect of sub-grid scale processes in coarse-resolution climate models if trained on high-resolution climate simulations. However, they may violate key physical constraints and lack the ability to generalize outside of their training set. Here, we show that physical constraints can be enforced in neural networks, either approximately by adapting the loss function or to within machine precision by adapting the architecture. As these physical constraints are insufficient to guarantee generalizability, we additionally propose to physically rescale the training and validation data to improve the ability of neural networks to generalize to unseen climates. 
\end{abstract}

\textbf{Code -} \url{github.com/tbeucler/CBRAIN-CAM}\\
\section{Introduction}
\label{sec:intro}

Computational resources limit climate models to coarse spatial
resolutions of order $10-100$km that distort convection and clouds \cite{Schneider2017}. This distortion causes well-known biases, including a lack of precipitation
extremes and an erroneous tropical wave spectrum \cite{Daleu2016}, which plague climate
predictions \cite{IPCC2014}. In contrast, global cloud-resolving models can resolve
scales of $\mathrm{O}\left(1\mathrm{km}\right)$, greatly reducing
these problematic biases. However, their increased computational cost
limits simulations to a few years. Machine-learning algorithms can
be helpful in this context, when trained on high-resolution simulations to replace semi-empirical
convective parametrizations in low-resolution models, hence bridging
the $1-100$km scales at reduced computational cost \cite{Gentine2018a,OGorman2018,Brenowitz2019}. In particular,
neural networks (NNs) are scalable and powerful non-linear regression
tools that can successfully mimic $\mathrm{O}\left(1\mathrm{km}\right)\ $ convective
processes (e.g., \cite{Brenowitz2018,Rasp2018}). However, NNs are typically physically-inconsistent by construction:
(1) they may violate important physical constraints, such as energy
conservation or the positive definition of precipitation, and (2)
they make large errors when evaluated outside of their training set,
e.g. produce unrealistically large convective heating in warmer climates.
In this conference paper, we ask:

\textit{How can we design physically-consistent NN models of convection?}

We adapt the NN's architecture of loss function to enforce conservation
laws in Section \ref{sec:Enforcing-Conservation-Laws} (review of \cite{Beucler2019a,Beucler2019b}) and improve the
NN's ability to generalize in Section \ref{sec:Improving-Generalization-to} by physically rescaling the data to transform extrapolation into interpolation without losing information (novel framework). 

In both sections, our ``truth'' is based on two years of simulations using
the Super-Parameterized Community Atmosphere Model version 3.0 \cite{Khairoutdinov2005} to simulate the climate in aquaplanet configuration \cite{Pritchard2014} with a realistic decrease in surface temperature from the Equator
to the poles. Snapshots of the interaction between climate- vs. convection-permitting scales are saved every 30 minutes,
which allows us to work in a data-rich limit \cite{Gentine2018a} by drawing 42M samples
from the first year for training and 42M samples from the second for
validation.

\section{Enforcing Conservation Laws in Neural Networks}
\label{sec:Enforcing-Conservation-Laws}

In this section, our goal is to conserve mass, energy and radiation
in a NN parametrization of convection. The parametrization's goal
is to map the local climate's state to the rate at which convection
redistributes heat and all three phases of water, along with radiation
and precipitation. In practice, we map a vector $\boldsymbol{x}\ $of
length 304 to a vector $\boldsymbol{y}\ $of length 218:\\
\begin{equation}
\boldsymbol{x}\overset{\mathrm{NN}}{\mapsto}\boldsymbol{y},
\end{equation}
where $ \boldsymbol{x}\ $groups local thermodynamic variables:
\begin{equation}
\boldsymbol{x}=\begin{bmatrix}\boxed{\boldsymbol{q_{v}}} & \boldsymbol{q_{l}} & \boldsymbol{q_{i}} & \boxed{\boldsymbol{T}} & \boldsymbol{v} & \boldsymbol{LS} & \boxed{p_{s}} & \boxed{S_{0}} & \boxed{\mathrm{SHF}} & \boxed{\mathrm{LHF}}\end{bmatrix}^{T},
\end{equation}
$ \boldsymbol{y}\ $groups subgrid-scale thermodynamic tendencies:
\begin{equation}
\begin{aligned}\boldsymbol{y}= & [ & \boxed{\boldsymbol{\dot{q}_{v}}} \ \ & \boldsymbol{\dot{q}_{l}} & \boldsymbol{\dot{q}_{i}}\ \  & \boxed{\boldsymbol{\dot{T}}} & \boldsymbol{\dot{T}_{\mathrm{KE}}}\ \  & \boldsymbol{\mathrm{lw}} & \boldsymbol{\mathrm{sw}}\\
 &  & \mathrm{LW_{s}}\ \  & \mathrm{LW_{t}} & \mathrm{SW_{s}}\ \  & \mathrm{SW_{t}} & P\ \  & P_{i} & ]^{T},
\end{aligned}
\end{equation}
where all variables are defined in Table 1 of the Supplemental Information (SI), and boxes are defined in Section \ref{sec:Improving-Generalization-to}. In this case, conservation laws
can be written as linear constraints acting on both input and output
vectors: $\boldsymbol{C}\begin{bmatrix}\boldsymbol{x} & \boldsymbol{y}\end{bmatrix}^{T}=\boldsymbol{0}$,
where we define the constraints matrix $\boldsymbol{C}\ $of shape
$4\times\left(304+218\right)$ in Equation (12) of \cite{Beucler2019b}.

\begin{figure}[htb]
\begin{centering}
\includegraphics[width=1\columnwidth]{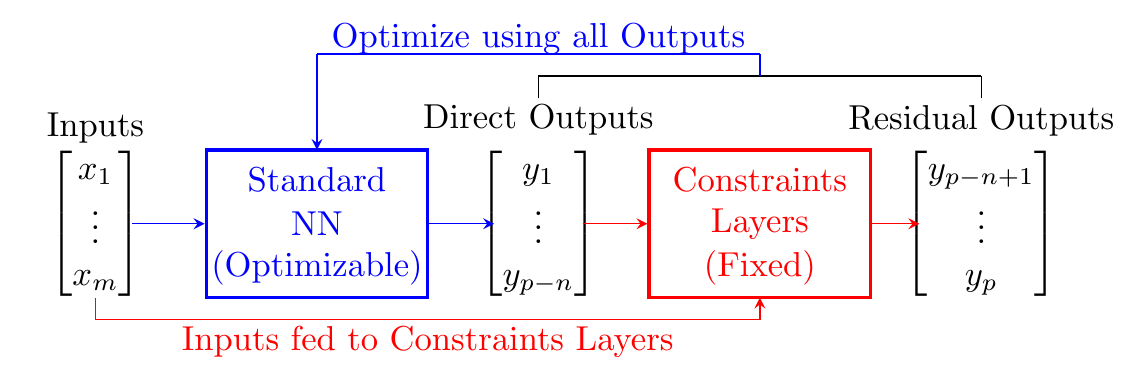}
\par\end{centering}
\caption{Architecture of ACnet\label{fig:Architecture-of-ACnets}}
\end{figure}

\begin{figure}[htb]
\begin{centering}
\includegraphics[width=1\columnwidth]{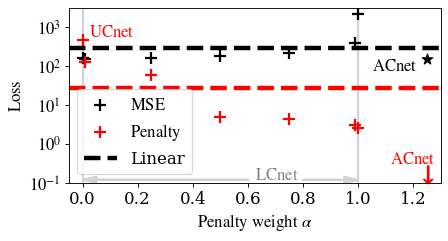}
\par\end{centering}
\caption{Mean-squared error $\mathrm{MSE}$ (black) and squared-residual $\mathrm{Penalty}$
(blue) from conservation laws for UCnet, ACnet and $\mathrm{LCnet}_{\alpha}\ $for
$\alpha\in\left\{ 0,0.01,0.25,0.5,0.75,0.99,1\right\} $. Note that
$\mathrm{LCnet}_{\alpha=0}=\mathrm{UCnet}$.\label{fig:Mean-squared-error-and}}

\end{figure}

\begin{figure}[htb]
\begin{centering}
\includegraphics[width=1\columnwidth]{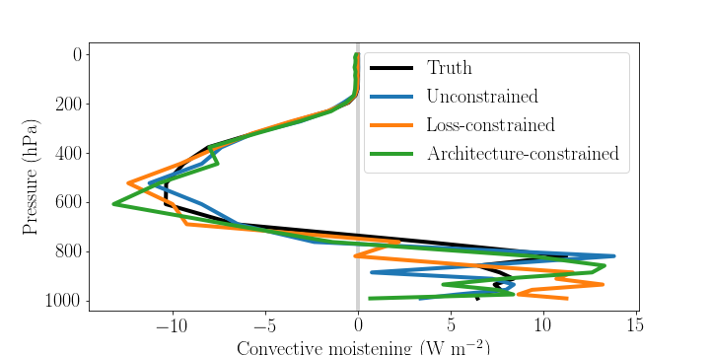}\\
\includegraphics[width=1\columnwidth]{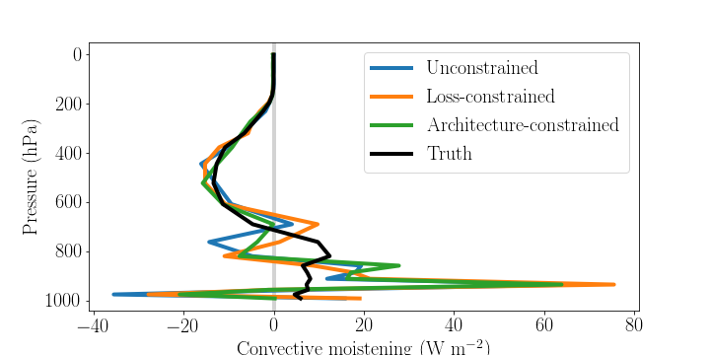}
\par\end{centering}
\caption{Daily-averaged prediction from UCnet, $\mathrm{LCnet_{\alpha=0.01}}$,
and ACnet in the Tropics for the reference climate (top) and the (+4K)
experiment (bottom)\label{fig:Daily-averaged-prediction-from}}

\end{figure}

We train a hierarchy of three NNs, all with a baseline architecture
of 5 layers with 512 neurons with optimized hyperparameters informed by a modern formal search using the SHERPA library \cite{Hertel2018}. All NNs are trained for 15 epochs using the RMSProp optimizer \cite{tieleman2012lecture}, which prevents over-fitting while guaranteeing reasonable performance. First, we train a baseline unconstrained
NN to map $\boldsymbol{x}\ $to $\boldsymbol{y}\ $(UCnet). Second,
we enforce conservation laws by introducing a penalty in the loss
function for violating conservation laws (similar to \cite{Karpatne2017,Jia2019}):
\begin{equation}
\mathrm{Loss}=\alpha\times\mathrm{Penalty}+\left(1-\alpha\right)\times\mathrm{MSE},\label{eq:Loss}
\end{equation}
where $\alpha\in[0,1]\ $is the penalty weight, wherein the penalty is given
by the mean squared-residual from the conservation laws:
\begin{equation}
\mathrm{Penalty}\overset{\mathrm{def}}{=}\frac{1}{304+218}\sum_{i=1}^{304+218}\left(\boldsymbol{C}\begin{bmatrix}\boldsymbol{x} & \boldsymbol{y}\end{bmatrix}^{T}\right)_{i}^{2},\label{eq:Penalty}
\end{equation}
and the mean-squared error is the mean squared-difference between
the NN's prediction $\boldsymbol{y_{\mathrm{pred}}}\ $and the truth
$\boldsymbol{y_{\mathrm{Truth}}}$:
\begin{equation}
\mathrm{MSE}\overset{\mathrm{def}}{=}\frac{1}{218}\sum_{i=1}^{218}\left(\boldsymbol{y_{\mathrm{Pred}}}-\boldsymbol{y_{\mathrm{Truth}}}\right)_{i}^{2}.\label{eq:MSE}
\end{equation}
The loss-constrained networks are referred to as $\mathrm{LCnet}_{\alpha}$.
Third, we enforce conservation laws by changing the NN's architecture
(see Figure \ref{fig:Architecture-of-ACnets}) so as to conserve mass,
energy and radiation to within machine precision \cite{Beucler2019a}. 
This NN, referred to as ACnet, calculates direct outputs using a standard
NN while the remaining outputs are calculated as residuals from the
fixed constraints layers, upstream of the optimizer. We summarize the $\mathrm{MSE\ }$and $\mathrm{Penalty}\ $of
UCnet, ACnet, and $\mathrm{LCnet}_{\alpha}\ $of various weights $\alpha$
in Figure \ref{fig:Mean-squared-error-and}. 
 In this weak constraint perspective, we note a clear trade-off between performance (measured by $\mathrm{MSE}$)
and physical constraints (measured by the $\mathrm{Penalty}$) as
the weight $\alpha\ $given to conservation laws increases from 0
to 1. An intermediate value is desirable (e.g., $\alpha=0.01\ $or
$\alpha=0.25$) as UCnet ($\mathrm{LCnet}_{\alpha=0}$) violates conservation
laws more than the multiple-linear regression baseline (horizontal
blue line) and $\mathrm{LCnet}_{\alpha=1}$ performs worse than the
multiple-linear regression baseline (horizontal black line). In contrast,
ACnet eliminates the need to compromise between performance and physical
constraints by enforcing conservation laws to within machine precision, which
is required in climate models, while achieving skill that is competitive with UCnet. Having solved the conservation issue, in the next section we turn to a deeper problem with NNs that, despite being physically-constrained,
ACnets and LCnets both fail to generalize to unseen climates (Figure \ref{fig:Daily-averaged-prediction-from}).

\section{Improving Generalization to Unseen Climates}
\label{sec:Improving-Generalization-to}

Testing generalization ability requires a generalization
test. For that purpose, we run another SPCAM simulation after
applying a uniform 4K warming to the surface temperature, which we
will refer to as the (+4K) experiment. We then test the NNs trained
on the reference climate in out-of-sample conditions, i.e. the deep
Tropics of (+4K) as illustrated in SI Figure 5. As can be seen in Figure \ref{fig:Daily-averaged-prediction-from},
NNs make extremely large errors when evaluated outside of their training
set, such as overestimating convective moistening by a factor of 5.

Motivated by the success of non-dimensionalization to improve the
generalizability of models in fluid mechanics, we seek to rephrase
the boxed part of our convective parametrization ($\boxed{\boldsymbol{x}}\mapsto\boxed{\boldsymbol{y}}$) to improve its generalizability by using variables that have similar distributions in both climates. Unlike idealized problems
in fluid mechanics, moist thermodynamics involve multiple non-linear
processes, including phase changes and non-local interactions, which prevent reducing our mapping to a few dimensionless numbers. Instead, we develop a three-step method
that consists of (1) non-dimensionalizing the input and output training
datasets to then (2) train NNs on these new datasets to finally (3)
compare their generalization abilities to our baseline UCnet. We use a different architecture of 7 layers with 128 neurons each for that lower-dimensional mapping, again informed by formal hyperparameter tuning, and train the NNs for 15 epochs using the Adam optimizer \cite{Kingma2014} while saving the state of best validation loss to avoid over-fitting. 

\begin{figure}[htb]
\begin{centering}
\includegraphics[width=1\columnwidth]{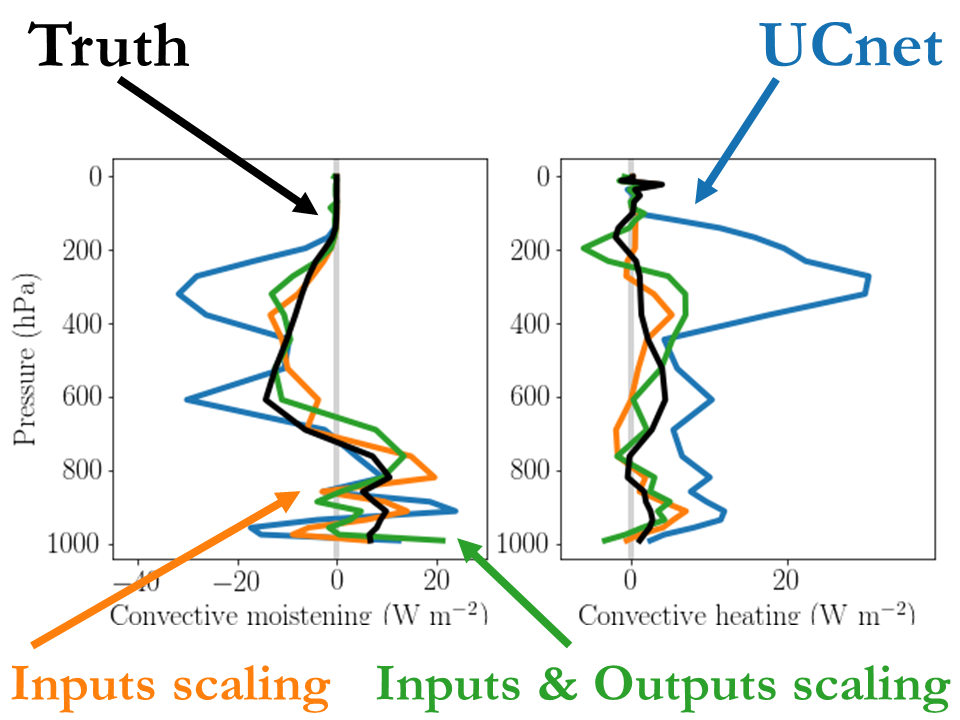}
\par\end{centering}
\caption{Daily-averaged predicted convective moistening (left) and heating
(right) from UCnet (blue), a NN with scaled inputs (orange line) and
a NN with scaled inputs and outputs (green) in the Tropics for the
(+4K) experiment\label{fig:Daily-averaged-predicted-convect}}

\end{figure}
We make progress via trial-and-error of multiple NNs and present two
successful rescalings below and in Figure \ref{fig:Daily-averaged-predicted-convect}: one for the inputs and one for the
outputs. 

Both successful rescalings leverage the Clausius-Clapeyron equation,
which implies that the saturation specific humidity $q_{v}^{*}\ $scales
exponentially with absolute temperature $T$, making the extrapolation
problem across climates especially challenging. The first rescaling (inputs) is
to use relative humidity instead of specific humidity (orange vertical
lines): $\boldsymbol{q_{v}}\mapsto\boldsymbol{RH}\approx\boldsymbol{q_{v}}/\boldsymbol{q_{v}^{*}}\left(T\right)$.
This exploits the fact that unlike specific humidity, relative humidity
is expected to change relatively little as the climate warms \cite{Romps2014}. The second rescaling (outputs) is to normalize the vertical redistribution
of energy by convection (in $\mathrm{W\ m^{-2}}$) using surface enthalpy
fluxes conditionally averaged on temperature (green line, see SI Equations 7 and 8). Although both physically-motivated rescalings significantly improve the ability of the NN to generalize to unseen
conditions, errors linked to the upwards shift of convection with warming are still visible in Figure \ref{fig:Daily-averaged-predicted-convect}.
This motivates physically-rescaling the vertical coordinate, which we leave for future work. 

Note that training a NN on both the reference climate and the (+4K)
experiment would be the simplest way of achieving generalizability.
However, the exercise of physically-rescaling the data to make our
NN generalize to unseen climates (1) mimicks the present-day challenge of predicting the future climate based on observed statistics of the current climate, and (2) identifies the relevant physical rescalings to work towards a climate-invariant mapping from the local climate to convective tendencies.

\section{Conclusion}
\label{sec:conclusion}

We made progress towards physically-consistent neural-network parametrizations
of convection in two ways. In Section \ref{sec:Enforcing-Conservation-Laws},
we enforced physical constraints in NNs (1) approximately
by using the loss function and (2) to within machine precision by modifying
the network's architecture. In Section \ref{sec:Improving-Generalization-to},
we helped neural-networks generalize to unseen conditions by leveraging
the Clausius-Clapeyron equation to physically rescale both inputs
and outputs of the parametrization. While our work initially stemmed
from operational requirements to improve convective processes in climate
models, the generalization exercise of Section \ref{sec:Improving-Generalization-to}
also offers a pathway towards data-driven scientific discovery of
the interaction between convection and the large-scale climate, e.g.
to better adapt the entrainment-detrainment paradigm to diverse convective
regimes or discover new equations to parameterize convection. 



\bibliographystyle{IEEEbib}
\bibliography{My_Collection.bib}

\pagebreak

\section*{Supplemental Information}

\begin{table}[htb]
\begin{tabular}{|c|c|}
\hline 
Variable & Name\tabularnewline
\hline 
$\mathrm{LHF}$ & Latent heat flux\tabularnewline
\hline 
$\mathrm{\boldsymbol{LS}}$ & Large-scale forcings in \tabularnewline
\ & water, temperature, velocity\tabularnewline
\hline 
$\boldsymbol{\mathrm{lw}}$ & Longwave heating rate profile\tabularnewline
\hline 
$\mathrm{LW_{s}}$ & Net surface longwave flux\tabularnewline
\hline 
$\mathrm{LW_{t}}$ & Net top-of-atmosphere longwave flux\tabularnewline
\hline 
$P$ & Total precipitation rate\tabularnewline
\hline 
$P_{i}$ & Solid precipitation rate\tabularnewline
\hline 
$p_{s}$ & Surface pressure\tabularnewline
\hline 
$S_{0}$ & Solar insolation\tabularnewline
\hline 
$\mathrm{SHF}$ & Sensible heat flux\tabularnewline
\hline 
$\boldsymbol{\mathrm{sw}}$ & Shortwave heating rate profile\tabularnewline
\hline 
$\mathrm{SW_{s}}$ & Net surface shortwave flux\tabularnewline
\hline 
$\mathrm{SW_{t}}$ & Net top-of-atmosphere shortwave flux\tabularnewline
\hline 
$\boldsymbol{T}$ & Absolute temperature profile\tabularnewline
\hline 
$\boldsymbol{\dot{T}}$ & Convective heating profile\tabularnewline
\hline 
$\boldsymbol{\dot{T}_{\mathrm{KE}}}$ & Heating from turbulent \tabularnewline
\ & kinetic energy dissipation \tabularnewline
\hline 
$\boldsymbol{q_{i}}$ & Ice concentration profile\tabularnewline
\hline 
$\boldsymbol{\dot{q_{i}}}$ & Convective ice tendency profile\tabularnewline
\hline 
$\boldsymbol{q_{l}}$ & Liquid water concentration profile\tabularnewline
\hline 
$\boldsymbol{\dot{q_{l}}}$ & Convective liquid water tendency profile\tabularnewline
\hline 
$\boldsymbol{q_{v}}$ & Specific humidity profile\tabularnewline
\hline 
$\boldsymbol{\dot{q_{v}}}$ & Convective water vapor tendency profile\tabularnewline
\hline 
$\boldsymbol{v}$ & North-South velocity profile\tabularnewline
\hline
\end{tabular}
\label{}
\caption{Definition of Variables: Variables that depend on height are (boldfaced) vectors, referred to as ``profiles''.\label{tab:Var_def}}
\end{table}

\begin{figure}[htb]
\begin{centering}
\includegraphics[width=1\columnwidth]{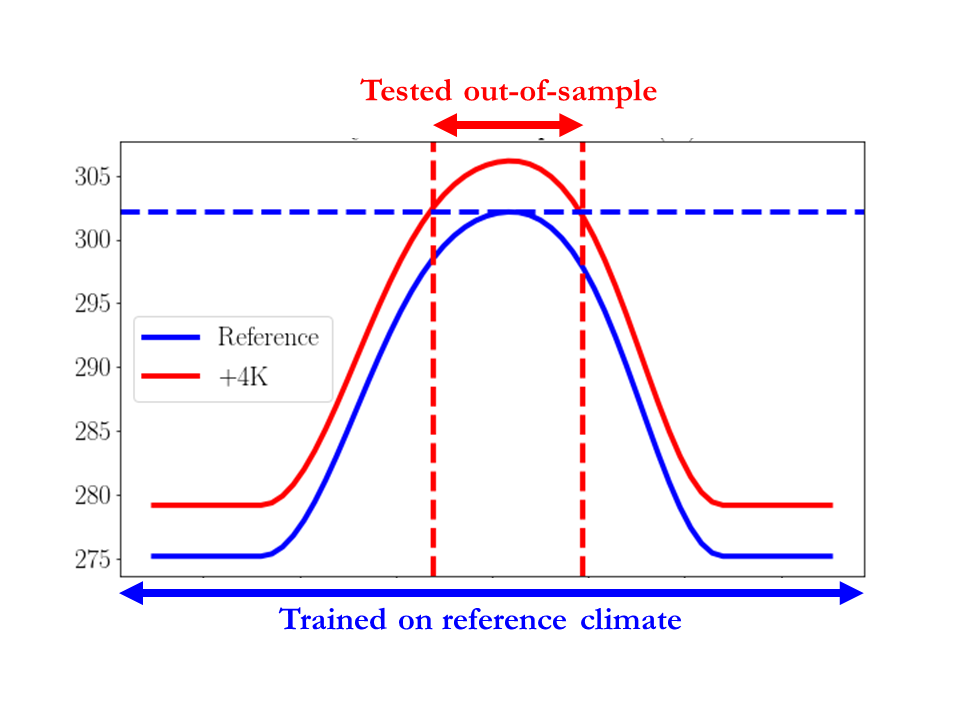}
\par\end{centering}
\caption{Surface temperature versus latitude for the reference and (+4K) experiments
\label{fig:Surface-temperature-versus}}
\end{figure}

\textbf{Physically-rescaling the NN outputs}\\

Motivated by the success of rescaling input data to take into account
the sharp increase of atmospheric water vapor concentration with temperature,
we develop an analogous normalization for the output data. As convection
typically redistributes energy from the bottom to the top of the atmosphere,
we can interpret convective heating and moistening profiles as processes
partitioning surface enthalpy fluxes (in $\mathrm{W\ m^{-2}}$) between
different layers of the atmosphere. As such, we mass-weight both profiles
before rescaling them using surface enthalpy fluxes conditionally-averaged
on near-surface temperature (referred to as $\mathrm{SEF\left(T_{NS}\right)}$).
Mathematically, this physical rescaling can be written as:
\begin{equation}
\boldsymbol{\dot{q}_{v}}\left[\mathrm{kg/kg}/s\right]\mapsto\frac{L_{v}\boldsymbol{\Delta p}}{g\times\mathrm{SEF}\left(T_{\mathrm{NS}}\right)}\boldsymbol{\dot{q}_{v}}\ \left[1\right],\label{eq:Convective_moistening_normalization}
\end{equation}
\begin{equation}
\boldsymbol{\dot{T}}\left[K/s\right]\mapsto\frac{c_{p}\boldsymbol{\Delta p}}{g\times\mathrm{SEF}\left(T_{\mathrm{NS}}\right)}\boldsymbol{\dot{T}}\ \left[1\right],\label{eq:Convective_heating_normalization}
\end{equation}
where $\boldsymbol{\Delta p}\ $are the layers' pressure thicknesses,
$g\ $is the gravity constant, $c_{p}\ $is the specific heat capacity
of dry air at constant pressure, and $L_{v}\ $is the latent heat
of vaporization of water in standard atmospheric conditions.

\end{document}